\documentclass[aps,prb,preprint,superscriptaddress]{revtex4-2}
\usepackage{amsmath,amssymb}
\usepackage{graphicx}  % needed for figures
\usepackage{dcolumn}   % needed for some tables
\usepackage{bm}        % for math
 \usepackage{float}
\usepackage{amsfonts}
\usepackage{gensymb}
\usepackage{upgreek}
\usepackage{mathrsfs} 
 \usepackage{hyperref}

\hypersetup{
    colorlinks=true,
    linkcolor=blue,
    filecolor=magenta,      
    urlcolor=blue,
}
 
\usepackage{blindtext,tikz}
\usetikzlibrary{quantikz}
\usetikzlibrary{calc}

\begin{document}

% Use the \preprint command to place your local institutional report
% number in the upper righthand corner of the title page in preprint mode.
% Multiple \preprint commands are allowed.
% Use the 'preprintnumbers' class option to override journal defaults
% to display numbers if necessary
%\preprint{}

%Title of paper
\title{Variational Quantum PageRank}

% repeat the \author .. \affiliation  etc. as needed
% \email, \thanks, \homepage, \altaffiliation all apply to the current
% author. Explanatory text should go in the []'s, actual e-mail
% address or url should go in the {}'s for \email and \homepage.
% Please use the appropriate macro foreach each type of information

% \affiliation command applies to all authors since the last
% \affiliation command. The \affiliation command should follow the
% other information
% \affiliation can be followed by \email, \homepage, \thanks as well.
\author{Christopher Sims}
\affiliation{Elmore Family School of Electrical and Computer Engineering, Purdue University}
\email{Sims58@purdue.edu}
%\email[]{Your e-mail address}
%\homepage[]{Your web page}
%\thanks{}
%\altaffiliation{}

%Collaboration name if desired (requires use of superscriptaddress
%option in \documentclass). \noaffiliation is required (may also be
%used with the \author command).
%\collaboration can be followed by \email, \homepage, \thanks as well.
%\collaboration{}
%\noaffiliation

\date{\today}

\begin{abstract}
The PageRank algorithm is used to rank web pages by their importance. Since it's development, the PageRank algorithm is a critical and fundamental part of search engines today. PageRank is a graph based algorithm that ranks pages based on how many other pages link to them.  This work develops a variational quantum version of the PageRank algorithm and compares the performance of the two algorithms. It is found that quantum PageRank performs better at ranking websites than the normal PageRank algorithm. 
\end{abstract}

\maketitle

\section{Introduction}
The original PageRank (PR) algorithm was originally developed by Larry Page and Sergey Brin in 1996\cite{Page1}. The PR algorithm uses an iterative method to minimize the values of a vector multiplied by an adjacency matrix which defines a graph. The higher the PageRank, the important that web page is considered to be ranked. PR is one of the most widely used algorithms for search engines results, social networking, and recommender systems such as recommendations for videos or ads \cite{Gleich2015}.

Since the PageRank has been introduced there have been numerous iterative improvements to decease the computational costs of large matrices and increase the performance of ranking websites\cite{sharma2020systematic,chung2014brief}. The actual PageRank algorithm utilized is kept secret by many corporations which use an improved version of the algorithm. However, researchers have shown that sparse matrix methods lead to a decrease in computational cost\cite{Davis2011,Yuster2005,Bell08}. It has also been shown that matrix representations such as QR decomposition or singular value decomposition (SVD) are better able to represent the matrix while reducing computational cost \cite{Goodall1993}. It is likely that the current algorithms used are a combination of these systems\cite{Berkhin2005,Xing2004,Ishii2018}. 

With the advent of artificial intelligence, there is likely a shift to utilizing large neural networks for recommender systems as oppose to pure PR to generate the results. Even in these systems, a PR representation is still commonly used. PR representation of graphs will still persist even with new machine learning models \cite{Burke2011,Zhang2020}.

This work presents a Variational Quantum PageRank (VQPR) algorithm that utilizes quantum singular value decomposition (QSVD) in order to represent the PR algorithm on the quantum computer. VQPR is compared to normal PR utilizing the Richardson method for the minimization.

\section{Richardson PageRank Algorithm}

The Richardson method is an iterative algorithm that updates the PageRank vector at each step by taking a weighted average of the PageRank values of the pages that link to it. This method has been shown to converge more quickly than the original pagerank method. 

For an adjacency matrix $A_{ij}$ if there is a link from node $j$ to node $i$ and $a_{ij} = 0$. There is a probability matrix $P = \frac{A}{||A||_\infty}$. The Richardson method then minimizes the function (see supplementary for full pseudo algorithm)\cite{richardson2001intelligent,richardson2006beyond,richardson2002mining} :
\begin{equation}
x_{i+1} = \alpha P x_i + (1- \alpha) \mathbf{1}/n
\end{equation}
where $\alpha$ is a damping factor, typically set to 0.85, and $\mathbf{1}$ is a vector of ones and $n$ is the length of vector $x$. 

\section{Quantum Singular Value Decomposition}
Quantum Singular Value Decomposition (QSVD) is a quantum algorithm that generalizes the classical singular value decomposition (SVD) to the quantum setting. SVD is a fundamental tool in linear algebra and has many applications in machine learning, data analysis, and scientific computing. QSVD extends SVD to the space of quantum states, allowing for the efficient decomposition of quantum states into a set of singular values and corresponding orthonormal bases \cite{Wang2021,Gilyen2019,bravo2020quantum,rebentrost2018quantum}.

Given a quantum state $\left|\psi\right\rangle$ on a $n$-qubit register, QSVD computes the singular values and orthonormal bases of the state. The QSVD algorithm is based on a combination of quantum phase estimation and quantum singular value transformation, which are quantum analogs of classical algorithms for computing eigenvalues and singular values, respectively\cite{miszczak2011singular,rebentrost2018quantum,wang2021quantum,wang2019quantum,gyongyosi2010quantum}.

To illustrate the QSVD algorithm, consider the example of a $2\times 2$ matrix $A = \begin{pmatrix} 1 & 0 \\ 0 & 2 \end{pmatrix}$.$A$ can be represented as a quantum state $\left|A\right\rangle$ on a $2$-qubit register by defining 
\begin{equation}
|A\rangle = \frac{1}{\sqrt{5}} (\sqrt{2} |00\rangle + |01\rangle + \sqrt{2} |10\rangle).
\end{equation}
QSVD is then applied to $|A\rangle$ to obtain the singular values and orthonormal bases of $A$. In this case, the singular values of $A$ are $\sigma_1 = 2$ and $\sigma_2 = 1$, and the corresponding orthonormal bases are
\begin{equation}
|u_1\rangle = \frac{1}{\sqrt{2}} (|00\rangle - |10\rangle), \quad |u_2\rangle = \frac{1}{\sqrt{2}} (|00\rangle + |10\rangle).
\end{equation}
The QSVD method can be seen as a generalized version of the (Variational Quantum Eigensolver) VQE method to non-Hermitian matrices \cite{Kandala2017,Wang2019,Tilly2022}.
\subsection{Richardson Variational Form of QVSD}
The QSVD algorithm allows for the efficient computation these singular values. As oppose to the method which calculates the singular values in the computational basis. The singular values that result from QSVD are kept in quantum space as hidden variables, thusly, the QSVD is never truly ``calculated''. The Richardson method for an iterative QSVD algorithm is defined as
\begin{equation}
x_{k+1} = ( \alpha P \cdot \sum_i^n M_i C_i ) + (1- \alpha) \mathbf{1}/n
\end{equation}

where $M_i$ is the probability to measure qubits in state $\ket{S_i}$ and $C_i $is the total number of counts per measurement after $n$ measurements. 

\subsection{Hardware Implementation}

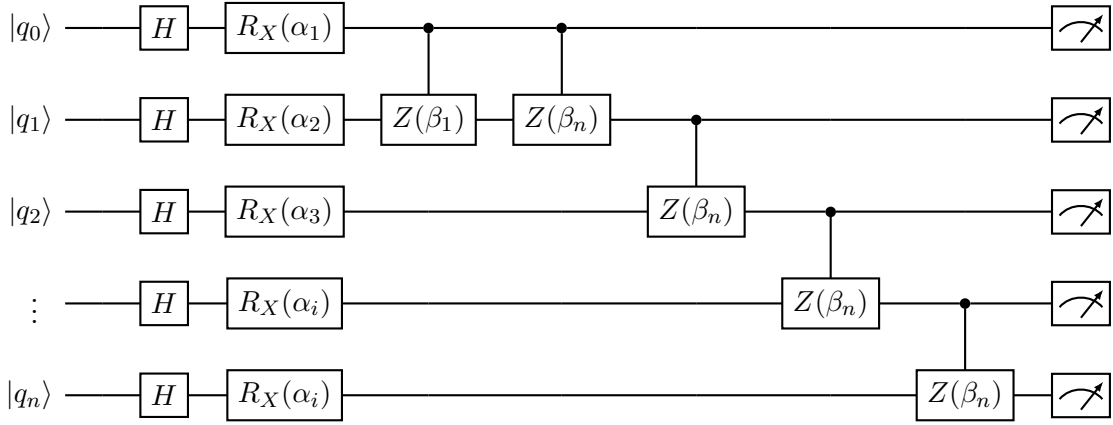
\begin{figure}[!ht]
\centering
\begin{quantikz}
 \lstick{\ket{q_0}} &\qw & \gate{H} &\gate{R_X(\alpha_1)}& \ctrl{1}& \ctrl{1}& \qw & \qw &\qw &\meter{}\\
\lstick{\ket{q_1}} &\qw & \gate{H} &\gate{R_X(\alpha_2)}&\gate{Z(\beta_1)}  &\gate{Z(\beta_n)} & \ctrl{1} & \qw &\qw &\meter{}\\
\lstick{\ket{q_2}} &\qw & \gate{H} &\gate{R_X(\alpha_3)} & \qw & \qw&\gate{Z(\beta_n)}&\ctrl{1}  &\qw &\meter{} \\
\lstick{$\vdots$\enspace} &\qw & \gate{H} &\gate{R_X(\alpha_i)} &\qw& \qw&\qw&\gate{Z(\beta_n)} &\ctrl{1}& \meter{}\\
\lstick{\ket{q_n}} &\qw & \gate{H} &\gate{R_X(\alpha_i)} & \qw&\qw&\qw&\qw&\gate{Z(\beta_n)} &\meter{}\\
\end{quantikz}
\caption{\textbf{QSVD}: An example of the QSVD algorithm for vector of length $n$. $\alpha_i = 2\pi x_i $ where x is a normalized vector $||x||_2 = 1$ and $\beta_i = 2\sin^{-1}(\frac{1}{2^{i}})$. For brevity, only the first and last set of CZ gates are shown}
\label{QSVD}
\end{figure}
The QSVD algorithm can be implemented on a quantum computer by first encoding the state vector $x$ into the quantum computer in the superposition state via apply an H-gate then a phased X-gate where each qubit corresponds to a component of the basis vector. For the main QSVD algorithm, phased CZ-gates are phased by $\beta_i$ applied in successive order up to $i+1$-qubits. This process is repeated for $n$-qubits [Figure \ref{QSVD}].
\section{Results and Discussion}
In order to analyze the performance of the Variational Quantum PageRank Algorithm (VQPR) an example $6\times6$ adjacency matrix ($A_{ij}$) is constructed:
\begin{equation}
\renewcommand{\arraystretch}{0.7}
A_{ij}= \begin{pmatrix}
0 & 1 & 1 & 0 & 0 & 0 \\
0 & 0 & 0 & 1 & 1 & 1 \\
0 & 0 & 0 & 0 & 1 & 1 \\
0 & 0 & 0 & 0 & 0 & 1 \\
0 & 1 & 0 & 0 & 0 & 1 \\
1 & 1 & 1 & 1 & 1 & 0 \\
\end{pmatrix}
\end{equation}
From $A_{ij}$ the transition probability matrix is constructed $P$, this matrix is utilized in the chardonnay method of both algorithms. For the VQPR, the convergence criteria is set to $\epsilon_{quantum} = 1\times 10^{-3}$ and for normal Richardson $\epsilon_{classical} = 1\times 10^{-6}$. For the classical Richardson method and the VQPR the resulting PageRanks are:
\begin{equation}
\begin{aligned}
X_{classical} &= [0.156, 0.170, 0.150, 0.135, 0.152, 0.240]\\
X_{quantum} &= [0.168, 0.087, 0.087, 0.087, 0.128, 0.443] 
\end{aligned}
\end{equation}
The quantum method converged after 5 iterations, while the classical method converged after 10 iterations. If the one norm is used for the probability matrix ($P = \frac{A}{||A||_1}$). The results of the PageRank are:
\begin{equation}
\begin{aligned}
X_{classical} &= [0.167, 0.167, 0.167, 0.167, 0.167, 0.167]\\
X_{quantum} &=[0.222, 0.121, 0.121, 0.121, 0.178, 0.238] 
\end{aligned}
\end{equation}
In this case, the quantum algorithm converged after 7 iterations while the classical algorithm fails to converge due to the matrix being nearly singular. 

In the $P_\infty$ form, the quantum gives node 6 a much high rating than the other nodes since it is much more connected, as does the classical algorithm. However, the quantum algorithm gives a very low rating to nodes [2,3,4] since they have low connectivity. In addition, the node 1 gets a very high rating because node 6 points to it but it does not point to node 6. In the classical algorithm, node [1,2,3,4,5] get similar ratings, due to their having 2 or 3 edges. As oppose to the classical algorithm, the matrix can also be represented by the $P_1$ form which can give a richer insight into the connectivity of the graph. In this representation, nodes 1 and 6 get a high rating due to their high connectivity to each other while the other nodes receive a low rating except node 5 which points to node 2 which also points back to it. 

While the performance of the quantum and classical algorithms are similar (with the assumption that each computational step is the same cost). This example shows that the VQPR algorithm is able to more richly discribe the connectivity between nodes because it is able to sample the entire phase space of the solution while the classical algorithm it an iterative solution.
\section{Methods}
\subsection{Simulation}
All calculations were performed in python v3.9.12 with the Google quantum AI Cirq package v1.0.0. GPU acceleration is enabled with the Nvidia cuQuantum 22.07.1 SDK. A qubits are simulated with a linear configuration with a grid layout similar to Google quantum machines. All qubits are measured in the $\hat{\mathrm{Z}}$ (computational or $\ket{0}$\&$\ket{1}$) basis. No noise is added in the calculation.
\subsection{Circuits}
The QSVD has two main parts, firstly, the state is encoded in the quantum register with H and phased RX gates in successive order for each element in vector $x$, the vector is assumed to be normalized so that the phases are between $0$ and $2\pi$. The next part of the algorithm is the QSVD unitary, these gate is constructed by applying a phase $\theta_i$ to a two qubits CZ gate, this phased gate is applied to $i$ qubits, this algorithm is repeated for all $n$-qubits. This leads to a time complexity of $O(n^2)$. Although this is complex in time, the SVD is represented in quantum space and does not need to be calculated classically. By measuring the quantum state $k = \log_2(n)$ times, it is possible to find the most probable eigenvector which represents the space in the computational basis. The total time cost for the algorithm is $O(n)$ for the measurement and $O(n^2)$ to prepare the state. This leads to a time complexity of $O(n^2)$.

\section{Conclusion}
In conclusion, it is shown that it is possible to represent the PageRank algorithm on a quantum computer. This work shows that a quantum computer is able to calculate an iterative matrix method with similar computational costs where the variational quantum PageRank (VQPR) has similar computational time to the Richardson method of the PageRank algorithm. However, VQPR is able to reach a more descriptive minima which better describes the connectivity of the adjacency matrix including long range connectivity. In future work, a representation of the adjacency matrix could be implemented and minimized on a quantum computer, without the need for a classical optimization step.

\section{Code availability}
Code will be made publicly available under the GNU GPL v3.0 public license upon full publication.

% Specify following sections are appendices. Use \appendix* if there
% only one appendix.
%\appendix
%\section{}

% \begin{figure}[ht]
% \includegraphics[width=1.0\textwidth]{Z_FIG.pdf}%
% \caption{\textbf{Autocorrelation:} The autocorrelation measurement averaged over all qubits $\overline{\langle Z(0)Zi(t)\rangle}$ with $g=0.9$ red denotes the low disorder system (D = 1) and blue denotes the high disorder system (D = 3.14) (A) 1D DTC system (B) the Fourier transform of the autocorrelation for the 1D system $\mathscr{F}\overline{\langle Z(0)Zi(t)\rangle}$ (C) the 2D DTC system with the same parameters as the 1D system (D) the Fourier transform of the autocorrelation for the 2D system $\mathscr{F}\overline{\langle Z(0)Zi(t)\rangle}$}
%\label{AC}
% \end{figure}

% If you have acknowledgments, this puts in the proper section head.
 \begin{acknowledgments}
C.S. acknowledges the generous support from the GEM Fellowship and the Purdue Engineering Horace W. and Helen K. ASIRE Fellowship.
 
 Correspondence and requests for materials should be addressed to C.S.
 (Email: Sims58@Purdue.edu)
\end{acknowledgments}

% Create the reference section using BibTeX:
%%%\bibliography{REFQPR.bib}
%aipnum4-2.bst 2019-01-14 (MD) hand-edited version of apsrev4-1.bst
%Control: key (0)
%Control: author (8) initials jnrlst
%Control: editor formatted (1) identically to author
%Control: production of article title (-1) disabled
%Control: page (0) single
%Control: year (1) truncated
%Control: production of eprint (0) enabled
%

\end{document}